\documentclass{amsart}
\usepackage{amsaddr}

\RequirePackage{amsthm,amsmath,amsfonts,amssymb, float, graphicx, bm, bbm, subfig, multirow, booktabs, array, longtable}

\usepackage{color}
\usepackage[justification=centering]{caption}
\allowdisplaybreaks

\theoremstyle{plain}

\newtheorem{theorem}{Theorem}[section]
\newtheorem{lemma}[theorem]{Lemma}
\newtheorem{corollary}{Corollary}
\newtheorem{proposition}{Proposition}
\newtheorem{assumption}{Assumption}
\theoremstyle{definition}
\newtheorem{definition}[theorem]{Definition}

\newtheorem{remark}{Remark}

\title[A Graph-based Approach to Estimating the Number of Clusters]{A Graph-based Approach to Estimating the Number of Clusters in High-dimensional Settings}
\author{Yichuan Bai and Lynna Chu}
\email{ycbai@iastate.edu; lchu@iastate.edu; }
\address{Department of Statistics, Iowa State University}

\begin{document}

\begin{abstract}
We consider the problem of estimating the number of clusters ($k$) in a dataset.  We propose a non-parametric approach to the problem that utilizes similarity graphs to construct a robust statistic that effectively captures similarity information among observations. This graph-based statistic is applicable to datasets of any dimension, is computationally efficient to obtain, and can be paired with any clustering technique. Asymptotic theory is developed to establish the selection consistency of the proposed approach. Simulation studies demonstrate that the graph-based statistic outperforms existing methods for estimating $k$, especially in the high-dimensional setting. We illustrate its utility on an imaging dataset and an RNA-seq dataset.

\end{abstract}
\maketitle


\section{Introduction}

Clustering is a fundamental unsupervised learning technique and a critical component of many statistics and machine learning pipelines. Cluster analysis seeks to partition a set of observations into $k$ groups (or clusters) with similar properties. The literature on clustering methods is vast; a handful of widely used methods include $K$-means \cite{macqueen1967kmeans, lloyd1982kmeans}; hierarchical algorithms \cite{ward1963hierarchical, king1967step}; spectral clustering \cite{donath1973spectral}; and model-based approaches using the expectation and maximization (EM) algorithm \cite{dempster1977maximum}. Many of these clustering approaches require the number of groups $k$ to be pre-specified, which can be challenging in the absence of knowledge regarding the true number of groups. While practitioners can sometimes use domain knowledge to estimate $k$, in practice the number of clusters $k$ is often unknown and needs to be estimated directly from the data. This problem becomes increasingly challenging in high-dimensional settings, when the curse of dimensionality can obscure correct estimation of $k$. With the increasing prevalence of high-dimensional data, it is essential to develop robust and reliable methods for accurately estimating the number of clusters in such settings.

Most existing approaches to estimate $k$ use distance-based criteria and are model-free. For example, early works largely focused on comparing the within-cluster dispersion ($W_k$) and between-cluster dispersion ($B_k$), both of which are calculated from distance-based measures \cite{calinski1974ch, milligan1985traceW, rousseeuw1987silhouettes}. If the distance is chosen to be squared Euclidean distance, then for $n$ observations, $x_1, \dots, x_n$, the within-cluster dispersion is defined to be $W_k = \sum_{j = 1}^k\sum_{i, i^{\prime} \in C_j}(x_i - x_{i^{\prime}})(x_i - x_{i^{\prime}})^T$,  and between-cluster dispersion is defined to be $B_k = \sum_{j = 1}^kn_j(\bar{x}_j - \bar{x})(\bar{x}_j - \bar{x})^T$, where $C_1, \dots, C_k$ are indice sets of observations in the $k$ clusters, $\bar{x}_j$ is the cluster mean of the $j$th cluster, and $\bar{x}$ is the centroid of the entire dataset. More recent developments include the gap statistic \cite{tibshirani2001gap}, which compares the within-cluster dispersion $(W_k$) with its expectation under a null reference distribution, an extension of the gap statistic that incorporates weights \cite{yan2007ddwei}, and the jump statistic \cite{sugar2003jump} which seeks to minimize a Mahalanobis-type distance. Other methods utilize cross-validation to estimate cluster stability \cite{fu2020cdpred, wang2010cv} or employ other resampling strategies \cite{monti2003consensus, fang2012bootstrap, tibshirani2005cluster}. Among these existing approaches, some were developed for specific clustering methods and cannot be applied broadly. Others lack theoretical guarantees or involve computationally intensive resampling in the form of cross-validation or bootstrap.

Alternatively, model-based approaches have been proposed that optimize an information criterion. For example, popular methods employ a model selection procedure for Gaussian mixture models to determine the number of mixture components \cite{fraley1998GMM, biernacki2000icl, fraley2007bayesian, fraley2002model}. However, these model-based approaches depend on assumptions regarding the data distributions, which may be too restrictive for practical applications. More recently, a data augmentation method for general model-based clustering was proposed \cite{luo2022determine}. All these model-based methods require the estimation of distribution parameters and may encounter limitations in the high-dimensional setting. Spectral clustering algorithms provide a different perspective; the number of clusters is estimated by analyzing eigenvalue gaps \cite{ng2001spectral, john2020spectrum} or eigenvectors \cite{zelnik2004self, xiang2008spectral} of the data affinity matrix. These spectral approaches may encounter issues due to noisy or computationally unstable eigenvectors and often lack theoretical support for the analysis of eigenvalues and eigenvectors.

\subsection{Estimating $k$ in high-dimensions}

In the high-dimensional setting, when the dimension of the observation $d$ exceeds the sample size $n$, existing approaches to estimate the number of clusters, $k$, can face challenges. In particular, using distance to assess cluster dispersion becomes less effective. As the dimensionality increases, observations tend to become approximately equidistant and sparsely distributed, which distance-based approaches to discern meaningful similarities among observations. 

\begin{figure}[!t]%
\centering
\includegraphics[width=1\linewidth]{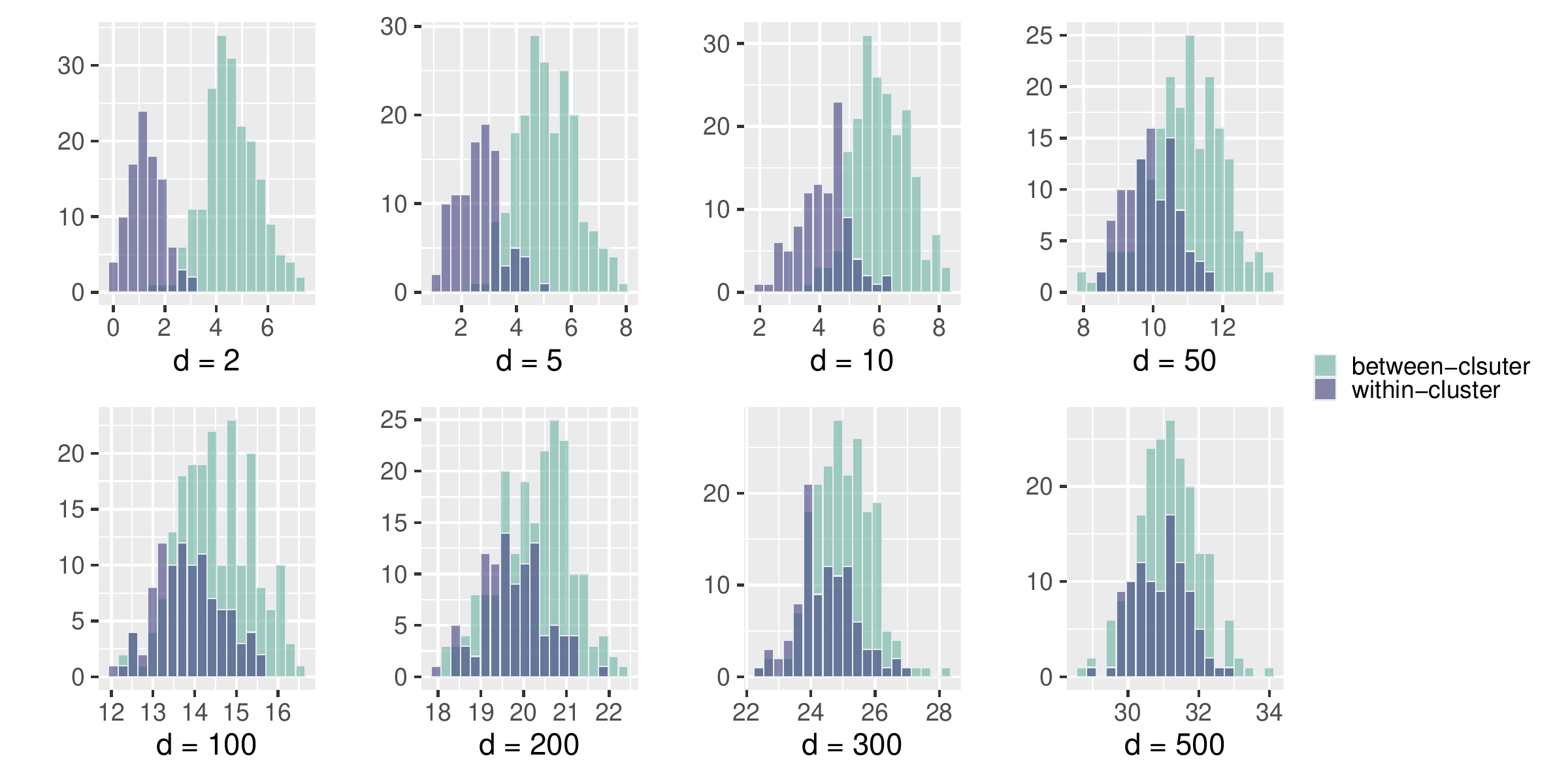}
\caption{Results for the three-cluster example: the histogram of the within-cluster distances and the between-cluster distances for dimensions 2, 5, 10, 50, 100, 200, 300, 500.}\label{fig:equidist}
\end{figure}

\begin{figure}[!t]%
\centering
\includegraphics[width=1\linewidth]{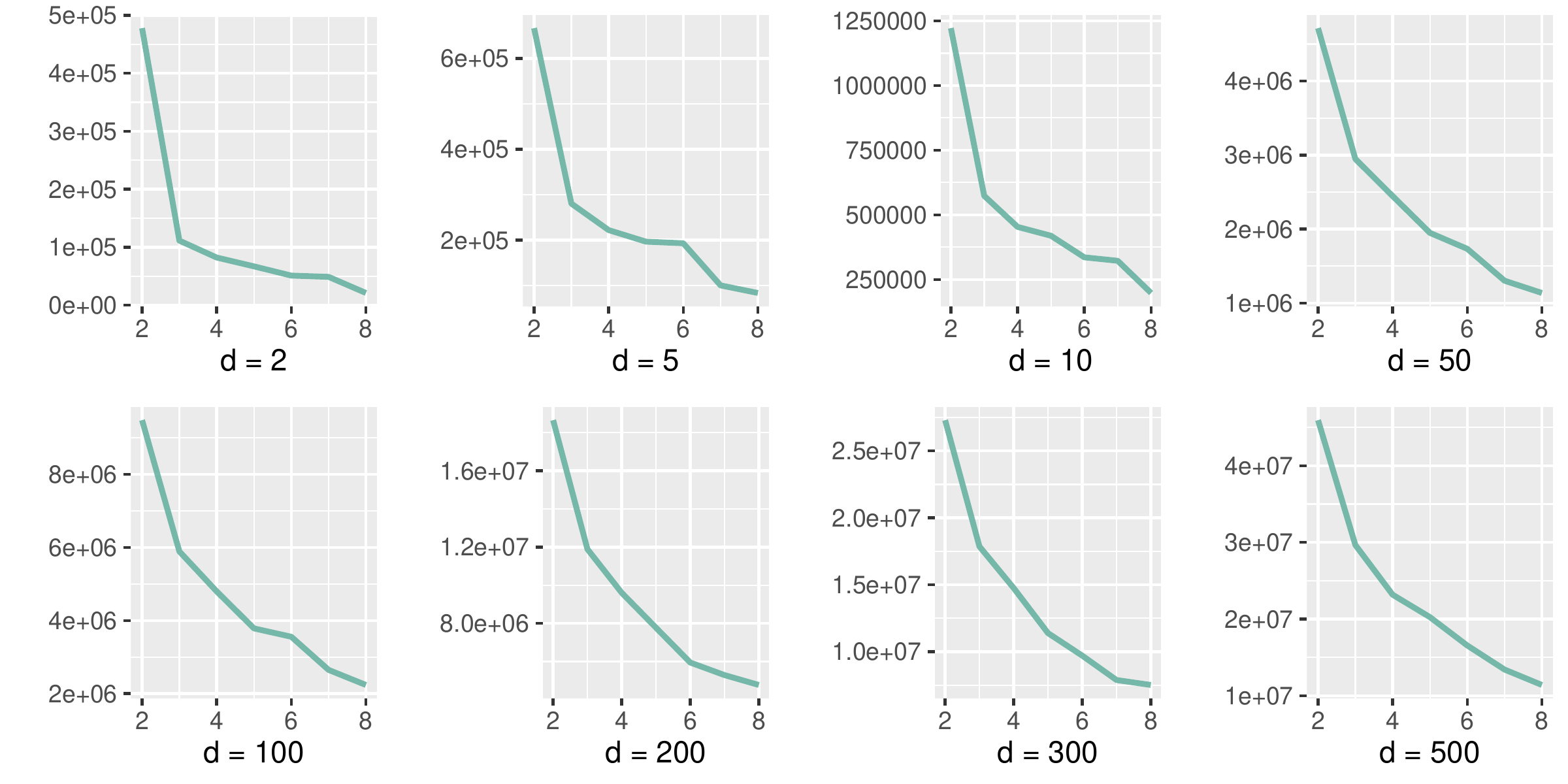}
\caption{Results for the three-cluster example: the within-cluster dispersion versus the number of clusters for dimensions 2, 5, 10, 50, 100, 200, 300, 500.}\label{fig:elbow}
\end{figure}

To illustrate the limitations of current distance-based approaches, we create an example dataset to show that using within-cluster dispersion ($W_k$) to estimate $k$ is ineffective for high dimensions. The example dataset has three clusters with 100 observations in each cluster. Observations are generated from Gaussian distributions that differ in mean: $x_1, \hdots, x_{100} \sim N_d(0, I_d)$, $x_{101}, \hdots, x_{200} \sim N_d(\sqrt{20/d}, I_d)$ and $x_{201}, \hdots, x_{300} \sim N_d(-\sqrt{20/d}, I_d)$. When the distance-based approaches are working properly, we would expect the (total) within-cluster dispersion to be smaller, on average, than the (total) between-cluster dispersion. From Figure \ref{fig:equidist}, we observe that when the dimension is relatively low ($d < 50$), this rationale holds. But as the dimension increases ($d \ge 50$), the distribution of within-cluster dispersion becomes almost identical to the between-cluster dispersion. Consequently, current approaches based on pairwise distances suffer severe limitations in their ability to distinguish within-cluster homogeneity in high dimensions. A common heuristic approach involves examining an elbow plot, which plots the within-cluster dispersion versus the number of clusters produced by a clustering procedure. Figure \ref{fig:elbow} shows such an elbow plot for various $d$. We can see that the change in the slope of the within-cluster dispersion before versus after the true number of clusters ($k = 3$) becomes less obvious as the dimension increases. This diminishing clarity makes the ad-hoc advice of looking for an `elbow' in the plot to choose the number of clusters unreliable. 

The tendency for observations to become nearly equidistant is caused solely by an increase in the dimensionality of data and is not related to the clustering assignments. In our illustration, since the clusters are well-separated, clustering accuracy under the true number of clusters is above 95\% across all settings. Accuracy is defined as the maximum proportion of correct labelings over all permutations of the clustering labels.

Model-based methods are also sensitive to an increase in dimensionality and can suffer from over-parameterization, along with difficulties in estimating parameters. An alternative approach is dimension reduction, where the dimension of the observations is reduced using techniques such as principal component analysis (PCA) or feature selection. However, these methods pose their own challenges, including the selection of principal components and incurring information loss. Additionally, many feature selection methods usually evaluate the data as a whole and may encounter difficulties if the informative dimensions of distinct clusters lie in different subspaces.

\subsection{Our Contribution}

We develop a non-parametric approach to estimate $k$ that can be applied to data in arbitrary dimensions, and is compatible alongside any clustering algorithm. Our approach can mitigate the curse of dimensionality, which may be problematic for purely distance-based approaches. Unlike model-based approaches, our approach does not require any distribution assumptions on the data or estimation of parameters. The key idea hinges on maximizing a statistic constructed from a similarity graph. We refer to this quantity as a graph-based statistic; it is a robust measure of the overall similarity among observations given their clustering assignment. This statistic stems from ideas first studied in the inference framework \cite{friedman_multivariate_1979, henze_multivariate_1988}, and more recently in \cite{chen_new_2017} and \cite{chen_weighted_2018}. We draw on similar ideas here and apply them to the problem of estimating $k$, the number of clusters. We demonstrate that our approach is especially useful when the dimension of observations is high. Our approach does not assume sparsity nor does it require any dimension reduction or feature selection step to work well. It is computationally straightforward, efficient to implement, and is shown to be highly effective in estimating $k$ in a broad range of settings. We establish the asymptotic selection consistency of our approach, which ensures that the estimated number of clusters ($\hat{k}$) converges in probability to the true number of clusters.

The paper unfolds as follows. Section \ref{sec:Method} introduces the graph-based statistic and our proposed approach to estimate the number of clusters. Section \ref{sec:consis} establishes the asymptotic properties of the proposed statistic. We present theoretical results that prove our method provides a consistent estimate of the true number of clusters. Simulation studies and real data applications are conducted in Section \ref{sec:sim} and Section \ref{sec:app}, respectively. We discuss the conditions under which our asymptotic theory holds and considerations on the density of the similarity graph in Section \ref{sec:dis}. Concluding remarks are given in Section \ref{sec:conclusion}.

\section{Graph-based estimation of $k$}\label{sec:Method}

\subsection{Problem setup}

Let $\{x_i\}, i = 1, \hdots, n$ be $d$-dimensional independent observations. Given an input $x\in \mathbb{R}^d $, let $\psi_k(\cdot)$ be a clustering assignment function, where $\mathbb{R}^d \rightarrow \{1, \hdots, k\}$ maps each observation in $\mathbb{R}^d$ to one of $k$ clusters. We use $\psi_k$ to denote the unique cluster assignment of $n$ observations into $k$ clusters generated by a given clustering algorithm. Given $\psi_k$, let $C_j, j = 1, \hdots, k$, denote the indices of the observations in cluster $j$, $n_j$ denote the number of observations in cluster $j$, and $p_j = n_j/n$ denote the proportion of observations in cluster $j$. We use $G$ to represent our similarity graph and use $|G|$ to represent the number of edges in $G$. Let $(u, v) \in G$ denote an edge in the graph $G$ that connects observation $x_u$ and $x_v$, $u\neq v \in \{1, 2, \dots, n\}$. Let $|G_i|$ denote the number of edges connected to observation $i=1, \hdots, n$ in similarity graph $G$.  

We aim to estimate the number of clusters that maximizes the separation of the observations across different values of $k>1$. The setting $k=1$ corresponds to no clustering, which we use as the null setting.

A definition of clustering is given in \cite{jain_data_2010}: 
\begin{quote} \small ``Given a representation of $n$ objects, find $k$ groups based on a measure of similarity such that the similarities between objects in the same group are high while the similarities between objects in different groups are low.''
\end{quote} 

There is no universal definition of similarity and, as such, no consensus on the definition of a cluster. Note that when the observations are not well separated, the notion of a cluster becomes difficult to define in the literature without domain knowledge. In this work, we assume that there is some (unknown) true clustering assignment, such that observations from the same density belong to the same cluster. This `true' number of clusters (denoted as $k^\star$) is defined based on the unique true clustering assignment presented in Definition \ref{def:optclus}.
\begin{definition}\label{def:optclus}
    Suppose the observations come from a set of $k^\star$ densities $\widetilde{\mathcal{F}} = \{f_j(x): j \in \{1\, \hdots, k^\star\},$ where $f_j, f_{j'}$ differ on a set of positive measures, $\forall j \neq j' \in \{1\, \hdots, k^\star\}\}$. The true clustering assignment is defined as $\widetilde{\psi}_{k^\star}: \mathbb{R}^p \rightarrow\{1,...,k^\star: X, Y \in C_j \text{ if and only if } X, Y \sim f_j(x), \forall j\in \{1,...,k^\star\}\}$, where $X$ and $Y$ are independent samples of observations. We refer to $k^\star$ as the true number of clusters under $\widetilde{\psi}_{k^\star}$. 
\end{definition}

Under true clustering, observations in the same cluster are from the same density, and observations in different clusters are from different densities. A difference in densities could imply either distinct distribution functions or the same distribution family but characterized by different parameter values. In practice, $f_j(x)$ is usually unknown and estimation about $k^\star$ relies only on observed data without any distributional assumptions regarding $f_j(x)$. Our definition of the clustering problem is a more general version of a mixture model \cite{fraley2007bayesian, luo2022determine}, as it imposes no assumption about the underlying distribution and parameters. 

\subsection{Graph-based statistic}
We propose a graph-based approach to estimate $k$. This approach utilizes a similarity graph that embeds observations into a graph structure, such that observations that are similar in some sense are more likely to have an edge connecting them. The similarity graph $G$ is constructed using all the observations $\{x_1, x_2, \hdots, x_n\}$, with each observation a node in the graph. Generally, the graph is constructed based on a similarity measure according to some criterion. For example,  $G$ could be a $\mathcal{K}$-minimum spanning tree ($\mathcal{K}$-MST) or a $\mathcal{K}$-nearest neighbor graph ($\mathcal{K}$-NN). A minimum spanning tree is a graph connecting all observations such that the sum of the distances across all edges is minimized. A $\mathcal{K}$-MST is a combination of $\mathcal{K}$ MSTs with disjoint edge sets, where the 2nd MST does not contain any edges in the 1st MST, the 3rd MST does not contain any edges in the first two MSTs, and so on. A $\mathcal{K}$-NN connects each observation to its $\mathcal{K}$ nearest neighbors. We assume that the similarity graphs are undirected. A standard choice for the similarity measure is Euclidean distance, but any informative similarity measure defined on the sample space can be used. A discussion on the choice of $\mathcal{K}$ is deferred to Section \ref{choi:k}.

Based on $G$, we calculate the graph-based statistic using quantities obtained from the graph. Specifically, we are interested in the within-cluster edge count, which we will denote as $R^G_{j}, j = 1, \hdots, k$. In what follows, we suppress the dependency on the graph in our notation and use $R_{j}$ for ease of readability. For a fixed $k$ and clustering assignment $\psi_k$,
           $$R_j = \sum_{(u, v)\in G}I(x_u \in C_j, x_v \in C_j),$$
where $j = 1, \dots, k$ and $I$ is the indicator function. Then, $R_j$ is the number of edges in the graph that connect observations within the same cluster $C_j$. 

If clustering is performed reasonably well, then the number of within-cluster edges should be relatively large since observations from the same cluster are more likely to connect to each other. Therefore, a relatively large $\sum_j^k R_j$ is indicative that the clustering assignment $\psi_k$ under $k$ sufficiently partitions similar observations together. 

\begin{figure}[!t]
\centering
\includegraphics[width=0.9\linewidth]{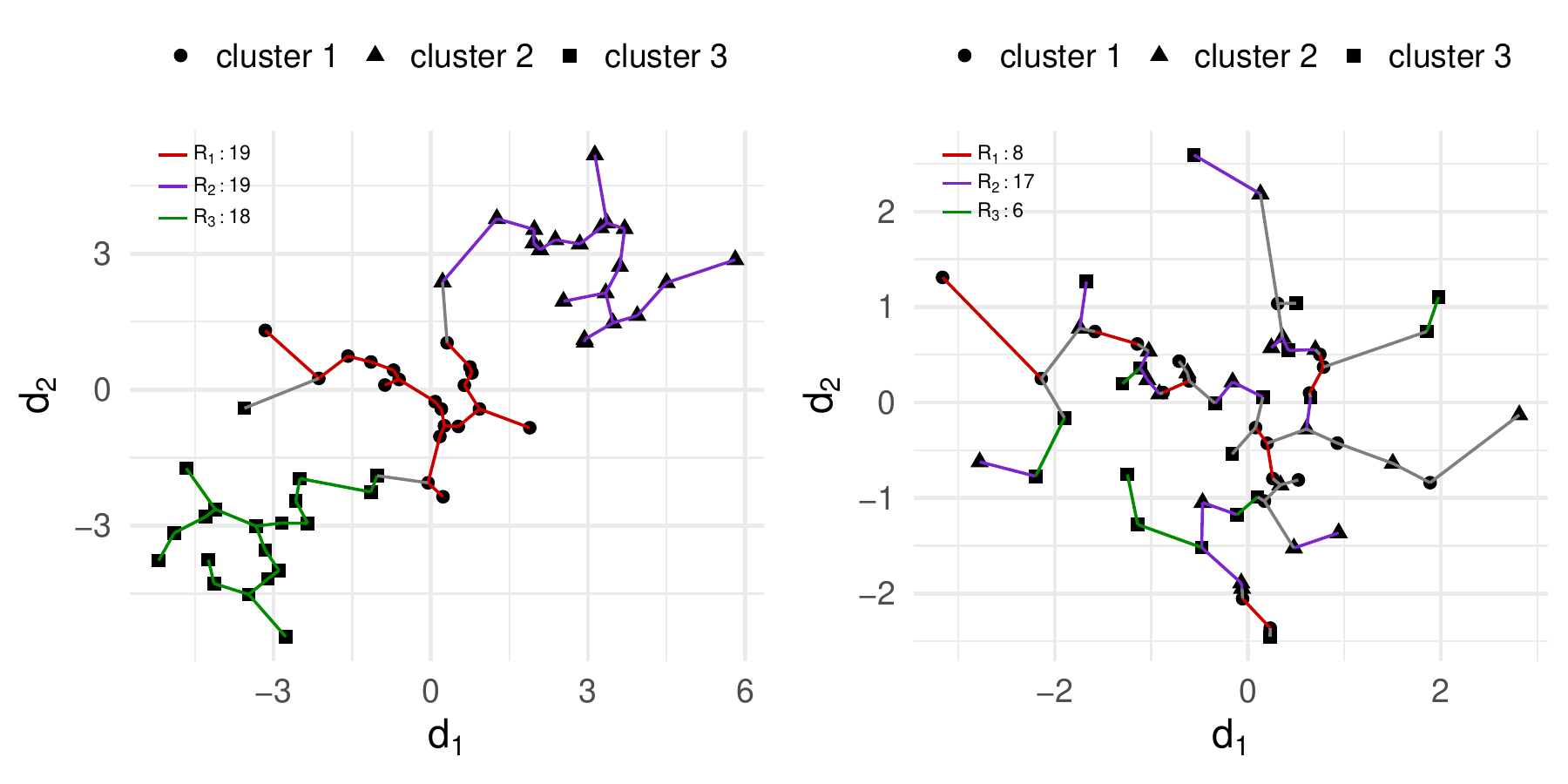}
\caption{Within-cluster edge-counts, denoted as $R_1$, $R_2$, $R_3$, count the number of edges connecting observations from the same cluster. Both similarity graphs are MSTs constructed from Euclidean distance. On the left, $x_i \in C_1 \sim \mathcal{N}(0,I_2)$, $x_i \in C_2 \sim \mathcal{N}((3,3)^T,I_2)$, $x_i \in C_3 \sim \mathcal{N}((-3,-3),I_2)$, for $i \in\{1, \dots, 60\}$ and $n_1 = n_2 = n_3 = 20$. On the right, all observations are drawn from $\mathcal{N}(0, I_2)$. }
\label{fig:exm_edge}
\end{figure}

To illustrate the calculation of $R_j$, we simulate a $d=2$ dimensional dataset consisting of three clusters $(k^\star = 3)$ with different centers, each containing 20 observations. An MST is constructed from the pooled observations and shown on the left-hand side of Figure \ref{fig:exm_edge}. The within-cluster edges are colored in red, purple, and green - these are edges that connect two observations from the same cluster. In this example, $R_1 = 19$, $R_2 = 19$, and $R_3 = 18$. 

When no clusters are present ($k^\star = 1$), the cluster labels are exchangeable. We define the null distribution of no clusters as the permutation distribution, where the $k$ cluster labels are randomly permuted.  The figure on the right in Figure \ref{fig:exm_edge} illustrates the within-cluster edges on a dataset under the null setting, which we generate by randomly permuting the three cluster assignment labels (maintaining $n_j$). We observe that the overall within-cluster edge counts ($R_1 = 8$, $R_2 = 17$, and $R_3 = 6$) generally decrease in comparison to the setting on the left. 

\begin{figure}[!t]
\centering
\includegraphics[width=0.9\linewidth]{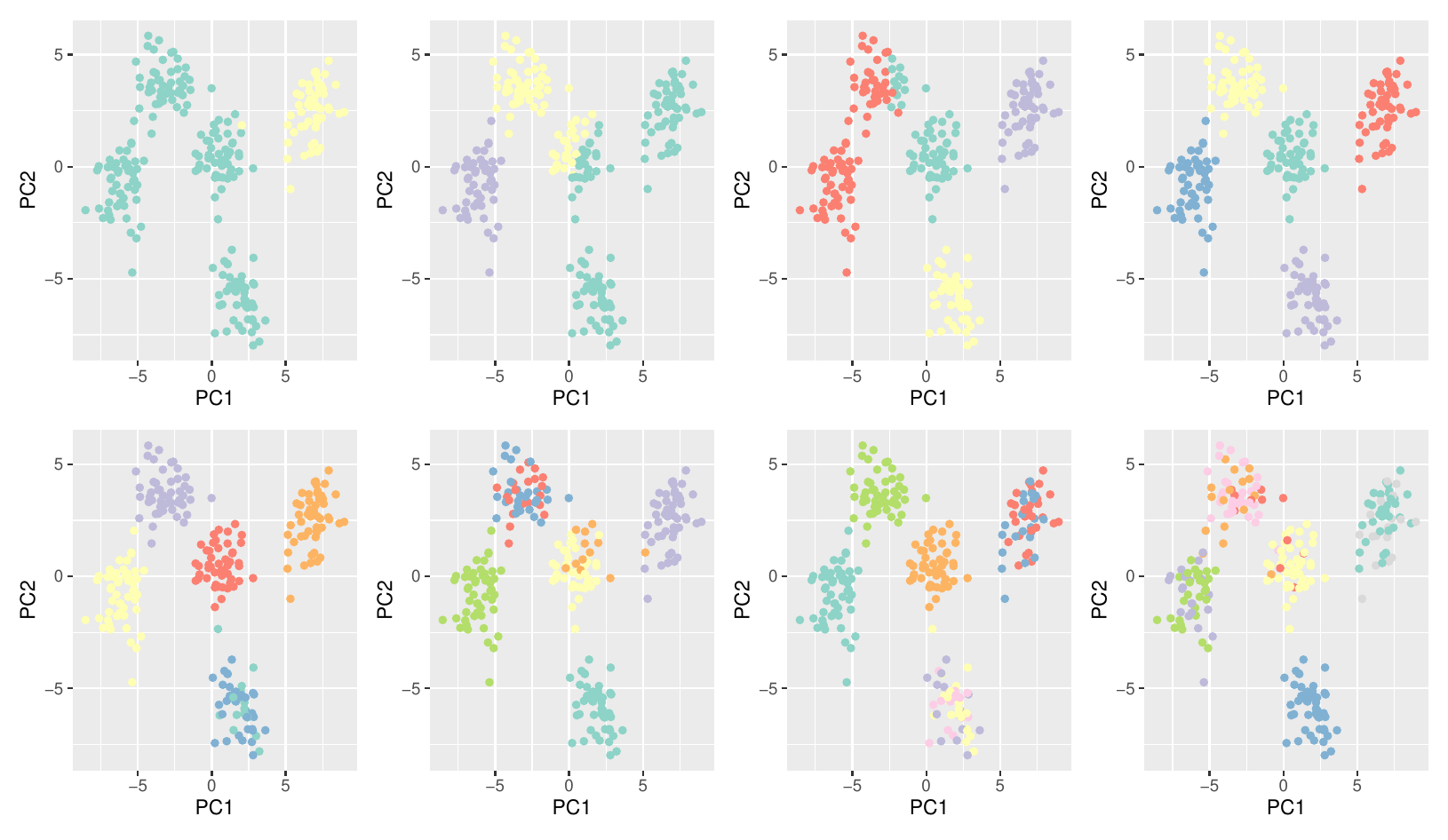}
\caption{Clustering assignment of the example dataset with $k^\star = 5$ for $k = $2, 3, 4, 5, 6, 7, 8, 9 (starting from left to right, top to bottom). Each plot shows the first two principal components of the data.}
\label{fig:clus}
\end{figure}

Under the null distribution, we derive the expectation and variance of the within-cluster edge counts (presented in Theorem \ref{th:asy_res}). An illustration of the behavior of $\sum_j^{k} R_j$  and $E(\sum_j^{k} R_j)$ are provided in Figure \ref{fig:clus}. An example dataset is generated with five clusters ($k^\star = 5$), each containing 50 observations with $d=200$; the means differ between clusters. For different $k$, the cluster assignment is shown in Figure \ref{fig:clus} such that different colors represent different cluster labels. 

\begin{figure}[!t]
\centering
\includegraphics[width=0.9\linewidth]{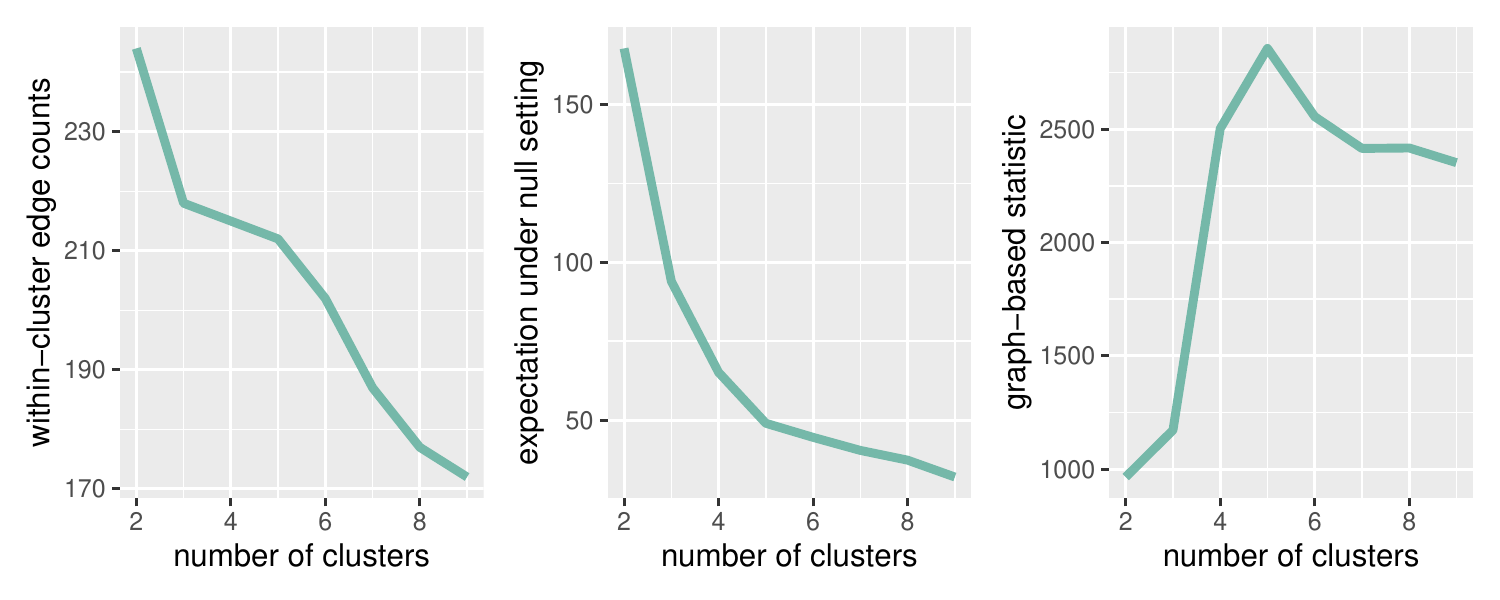}
\caption{Simulated dataset with $k^\star = 5$. Left: number of within-cluster edges ($\sum_{j=1}^{k} R_j$) versus $k$, middle: the expectation of the within-cluster edge counts under the null distribution ($E(\sum_{j=1}^{k} R_j$)) versus $k$, right: the graph-based statistic ($Q(k)$) versus $k$. }
\label{fig:inc_edge}
\end{figure}

As $k$ increases, the data are split into smaller groups and it becomes harder to form an edge within a cluster. We observe that  in Figure \ref{fig:inc_edge}, both $\sum_j^k R_j$ and $E(\sum_j^{k} R_j)$ decrease as $k$ increases. In order to make the within-cluster edge count comparable across different values of $k$, we standardize $\sum_j^{k} R_j$ by comparing it to its expectation and variance under the null distribution, and refer to the standardized statistic as the graph-based statistic, denoted as $Q(\psi_k)$. We see on the far right plot of Figure \ref{fig:inc_edge} that once properly standardized, the graph-based statistic is maximized at the true number of clusters ($k^\star =5$).

Our graph-based statistic is defined as follows:
\begin{equation}
    Q(\psi_k) = \frac{\left(\sum_{j = 1}^k\frac{1}{n_{j}}R_{j} - E(\sum_{j = 1}^k\frac{1}{n_{j}}R_{j})\right)^2}{\text{Var}\left(\sum_{j = 1}^k\frac{1}{n_{j}}R_{j}\right)}.
\end{equation}

Let $k_{max}$ be the largest number of clusters considered.  We evaluate the graph-based statistic for each value of $k$ from $2$ to $k_{max}$ and estimate the number of clusters, $\hat{k}$, to be: 
 \begin{equation}
 \hat{k} := \arg\max_{2\leq k \leq k_{max}} Q(\psi_k) .
 \label{eq:maxk}
 \end{equation}
 
Note that, to calculate the edge counts $\sum_{j = 1}^k\frac{1}{n_{j}} R_j$ for various values of $k$, the similarity graph only needs to be generated once. The graph depends solely on the values of $x_i, i = 1, \hdots, n$ and not the cluster assignments themselves. Since we derive explicit analytical expressions for the expectation and variance, this makes any direct resampling, such as permutation or bootstrap, unnecessary for calculating our proposed graph-based statistic. This alleviates a substantial computational burden found in other existing methods, making our approach competitively fast and efficient. The expressions for the expectation, variance, and covariance under the permutation null are presented in Theorem \ref{th:asy_res}.

\begin{theorem}\label{th:asy_res}
The expectation, variance, and covariance of the within-cluster edge count $R_j$, $j = 1, 2, \dots, k$ under the permutation null distribution are as follows:
\begin{align*}
 E(R_{j}) =& |G|\frac{n_j(n_j - 1 )}{n(n - 1)},\\
 \text{Var}(R_{j}) 
=&\frac{n_j(n_j - 1 )(n-n_j)(n-n_j - 1 )}{n(n - 1)(n - 2)(n - 3)}\left(|G|+\frac{n_j-2}{n-n_j-1}G_C - G_E\right),\\
\text{Cov}(R_{j}, R_{j'}) 
=&\frac{n_jn_{j'}(n_j - 1 )(n_{j'} - 1 )}{n(n - 1)(n - 2)(n - 3)}\left(|G|- G_C - G_E\right),
\end{align*}
where $G_C = \sum_{i = 1}^n|G_i|^2 - \frac{4}{n}|G|^2$, $G_E = \frac{2}{n(n-1)}|G|^2$, and $G_i$ is the set of edges in $G$ that connects to observation $x_i$.
\end{theorem}

The proof of Theorem \ref{th:asy_res} is given in Supplement Section S2. The expectation and variance of the sum of the within-cluster edge counts are linear combinations of the expectation, variance and covariance of the within-cluster edge counts.

\begin{corollary}
Under the permutation null distribution,
\begin{align*} 
E\left(\sum_{j = 1}^k\frac{1}{n_{j}}R_{j}\right) & = \sum_{j = 1}^k\frac{1}{n_{j}}E(R_{j}),\\
\text{Var}\left(\sum_{j = 1}^k\frac{1}{n_{j}}R_{j}\right) & = \sum_{j = 1}^k\frac{1}{n_{j}^2}\text{Var}(R_{j}) + 2\sum_{j =1 }^k\sum_{j'>j }^k\frac{1}{n_{j}n_{j'}}\text{Cov}(R_{j}, R_{j'}).
\end{align*} 
\end{corollary}

\section{Selection consistency}\label{sec:consis}

We establish the asymptotic consistency of the graph-based statistic. Our results show that the estimator $\hat{k}$ converges in probability to the true number of clusters $k^\star$ under $\widetilde{\psi}_{k^\star}$, assuming the divergence in clusters containing a mixture of densities has a lower bound. 

\subsection{Asymptotic properties of graph-based quantities}

We first present some asymptotic properties of the graph-based quantities $R_j$, $E(R_j)$, $\text{Var}(R_j)$, $\text{Cov}(R_{j}, R_{j'})$, $\forall j \neq j' \in \{1, \dots, k\}$. 

\begin{theorem}\label{th:exist_lim}
If the similarity graph is constructed using $\mathcal{K}$-MST or $\mathcal{K}$-NN, where $\mathcal{K} = O(1)$, as $n\rightarrow\infty$ with $n_j/n \rightarrow p_j$ for $j = 1, 2, \dots, k$, then under the permutation null distribution, we have
\begin{align*}
&\lim_{n\rightarrow\infty}\frac{1}{n}\left(\sum_{i = 1}^n|G_i|^2 - \frac{4}{n}|G|^2\right)= \mathcal{B} < \infty,\\
&\lim_{n\rightarrow\infty}\frac{E(R_{j})}{n} = \mathcal{K}p_j^2,\\
&\lim_{n\rightarrow\infty}\frac{\text{Var}(R_{j})}{n} = p_j^2(1-p_j)^2(\mathcal{K} + \frac{p_j}{1-p_j}\mathcal{B}),\\  
&\lim_{n\rightarrow\infty}\frac{\text{Cov}(R_{j}, R_{j'})}{n} = p_j^2p_{j'}^2(\mathcal{K} - \mathcal{B}).
\end{align*}
\end{theorem}
The proof of Theorem \ref{th:exist_lim} is given in Supplement Section S3. 

\begin{lemma}\label{le:alterlimit}
     Let the density function in cluster $C_j$ be $f_j, j = 1, ...k$ with $f_j \neq f_{j'}$ for $j \neq j'$. If $n_j \rightarrow \infty$ with $n_j/n \rightarrow p_j$ for $j = 1, 2, \dots, k$, and if the similarity graph is constructed using $\mathcal{K}$-MST or $\mathcal{K}$-NN, where $\mathcal{K}=O(1)$. Then,
\begin{align}
\frac{R_{j}}{n} &\rightarrow \mathcal{K}\int\frac{p_j^2f_j^2(x)}{\sum_{s = 1}^kp_sf_s(x)}dx  \quad \text{almost surely for }j = 1, 2, \dots, k.
\end{align}

\end{lemma}

This result is an extension of the arguments made in \cite{henze1999kmst} and \cite{schilling1986knn} for $\mathcal{K}$-MST and $\mathcal{K}$-NN, respectively. The proof of Lemma \ref{le:alterlimit} is given in Supplement Section S4.

\subsection{Main result}
We define the limiting quantity of $Q(\psi_k)$ to be: 
\begin{align*}
\mathcal{I}(\psi_k)& =  \lim_{n\rightarrow\infty}\frac{Q(\psi_k)}{n}.
\end{align*}

\begin{proposition}\label{pro:exist}  As $n_j \rightarrow \infty$ with $n_j/n \rightarrow p_j$ for $j = 1, 2, \dots, k$,
    $\mathcal{I}(\psi_k)$ exists and is finite.
\end{proposition}

The proof of Proposition \ref{pro:exist} is in Supplement Section S5.

\begin{proposition}\label{pro:consismax}
For a finite set $\mathcal{A}$, if $f_n(x) \rightarrow f(x)$, $\forall x \in \mathcal{A}$, then for the maximizer $x^n = \arg\max_x f_n(x)$ and the unique maximizer $x^\star = \arg\max_x f(x)$, we have $x^n \rightarrow x^\star$ almost surely.
\end{proposition}

The proof of Proposition \ref{pro:consismax} is given in Supplement Section S6
. According to Proposition \ref{pro:consismax}, what remains to establish the consistency of our estimator $\hat{k}$ is to prove that the maximizer of $\mathcal{I}(\cdot)$ occurs under clustering assignment $\widetilde{\psi}_{k^\star}$ with $k^\star$ clusters. 

Given a clustering assignment $\psi_k$, we consider two paths to prove that $k^\star$ is the maximizer of $\mathcal{I}(\cdot)$. 
For a given clustering assignment, if any cluster contains a mixture of densities, we will refer to this as a mixture cluster. In this case, the quantity $\mathcal{I}(\cdot)$ can be increased by generating a new improved clustering assignment where the mixture cluster is split into homogenous subgroups. On the other hand, it is possible that for a given $\psi_k$, multiple clusters contain observations that share the same density. In that case, an improved clustering that combines these clusters also increases the quantity $\mathcal{I}(\cdot)$. Using these two steps, the maximum of the limitation is reached at the true number of clusters $k^\star$.

We define a criterion to measure the divergence between any two densities.

\begin{definition}
Given a finite set of probability density functions $\mathcal{F} = \{f_s$, $s \in Q$; such that  $f_s, f_{s'} \text{ differ on a set of positive measure, }\forall s \neq s' \in Q\}$, where $Q$ is finite, $p_s$ is a weight such that $p_s > 0$ and $\sum_{s\in Q} p_s = 1$, then the divergence between $f_j$ and $f_{j'}$, $j\neq j' \in Q$ is defined as:
\begin{equation*}
    \mathcal{D}(f_j, f_{j'}|\mathcal{F}) = \int\frac{p_{j}p_{j'}(f_{j}-f_{j'})^2 }{\sum_{s \in Q}p_sf_s}.
\end{equation*}
\end{definition}

\begin{assumption} \label{as:1}
Assume a true clustering exists according to Definition \ref{def:optclus}. Under the set of densities $\widetilde{\mathcal{F}}$, the total divergence of densities in all mixture clusters has a uniform lower divergence bound:

\begin{align}\label{eq:ass}
               \sum_{j \in H'}\frac{1}{p_{j}}\sum_{\substack{t,r=1, \\t \neq r}}^{m_j}\mathcal{D}(f_{j_t}, f_{j_r}|\widetilde{\mathcal{F}})
        >\left(1-\frac{\sqrt{k-1}}{\sqrt{k+k^+-1}}\right)\sum_{\substack{j, j' =1 \\j\neq j'}}^{k^\star} \mathcal{D}(f_{j}, f_{j'}|\widetilde{\mathcal{F}})
\end{align}
for a given a clustering assignment $\psi_k$. Here, $p_j = n_j/n$, $m_j$ is the number of densities in mixture cluster $j$, $H'$ is the index set of clusters containing mixtures for a clustering assignment $\psi_k$, $|H'|$ is the number of clusters that contain mixtures, and $k^+ = \sum_{j \in H'}m_j-|H'|$.
\end{assumption}

The assumption posits that the divergence between clusters in the mixture cluster must be bounded by a proportion of the total pairwise divergence under true clustering. This is a sufficient condition to establish Lemma \ref{lem2}. In order to achieve better performance by splitting mixtures into sub-groups and creating a new cluster assignment, the divergence in the mixtures must satisfy a lower bound. In practice, this assumption is difficult to validate since the true cluster densities are unknown. We provide some illustrative examples in Section \ref{dis:ass}, where the assumption can be verified given knowledge of the true cluster densities, and the corresponding integrals are relatively straightforward to compute.
\begin{lemma}\label{lem2}
       Under Assumption \ref{as:1}, when the cluster assignment $\psi_k$ generates mixture clusters consisting of different densities, let $S$ be the index set of clusters after splitting mixture clusters into homogeneous sub-clusters, and let the cluster assignment $\psi_l$ be defined as $\psi_l: R^p \rightarrow S$, where $l = |S|$. Then we have $\mathcal{I}(\psi_k) < \mathcal{I}(\psi_l).$
        \end{lemma}

The proof of Lemma \ref{lem2} is presented in Supplement Section S8.

Lemma \ref{lem1} establishes that when the clustering assignment contains many clusters with observations from the same density, the value of $\mathcal{I}(\cdot)$ can be increased by combining or merging these clusters. 

\begin{lemma}\label{lem1}

 For a clustering assignment $\psi_k$ ($k>k^{\star}$), if all clusters are homogenous and there are observations from multiple clusters following the same density in $\widetilde{\mathcal{F}}$, we have $\mathcal{I}(\psi_k) <\mathcal{I}(\widetilde{\psi}_{k^\star}).$
  
\end{lemma}

The proof of Lemma \ref{lem1} is given in Supplement Section S7. Lemma \ref{lem1} proves that the value of $\mathcal{I}(\cdot)$ will increase by combining the homogeneous cluster structures into the same cluster. Lemma \ref{lem1} assumes that all the mixture clusters have already been split and that observations within the remaining clusters belong to the same density. 

\begin{remark}[Uniqueness]  
There exists a unique cluster assignment $\widetilde{\psi}_{k^\star}$ with the true number of cluster $k^\star$ that maximizes $\mathcal{I}(\cdot)$.
\end{remark}

For any other clustering that is different than the true clustering, we can use Lemma \ref{lem2} and Lemma \ref{lem1} to prove that the quantity $\mathcal{I}(\cdot)$ under such clustering is smaller than $\mathcal{I}(\widetilde{\psi}_{k^\star})$, such that the cluster assignment $\widetilde{\psi}_{k^\star}$ is unique. 

\begin{theorem}\label{th:main}
Suppose Assumption \ref{as:1} holds and a true clustering exists. Let $\hat{k}$ be the graph-based estimator $\hat{k}$ as in (\ref{eq:maxk}). Then,
    $$Pr(\hat{k} = k^\star) \rightarrow 1 \text{ if } n \rightarrow \infty \text{ and } n_j/n \rightarrow p_j \in (0, 1).$$
\end{theorem}

\begin{proof}[Proof.\nopunct] 
    Recall that $\hat{k}$ is the maximizer of the graph-based statistic as defined in Equation \ref{eq:maxk}. By Lemma \ref{lem2} and Lemma \ref{lem1}, we establish that $\widetilde{\psi}_{k^\star}$ with $k^\star$ clusters is the maximizer of the quantity $\mathcal{I}(\cdot)$, which is the limit of the graph-based statistic as $n \rightarrow \infty$.  Then, by Proposition \ref{pro:consismax}, $\hat{k} \rightarrow k^\star$ almost surely and the estimator is consistent.
\end{proof}

We establish that the limiting quantity $\mathcal{I}(\cdot)$ is maximized at the true number of clusters $k^\star$ under $\widetilde{\psi}_{k^\star}$ ; this allows us to prove the selection consistency of our method, which gives validity to utilizing the graph-based estimator as an estimate for $k$. We demonstrate in Section \ref{sec:sim} that, even for finite samples, our method can still accurately recover the true number of clusters in high-dimensional settings. 

\section{Simulation Studies}\label{sec:sim}
In this section, we present several simulation studies to evaluate the performance of the proposed method. We compare our method to the following commonly used methods to estimate the number of clusters: the TraceW statistic \cite{milligan1985traceW}; the Silhouette statistic \cite{rousseeuw1987silhouettes}; the Gap statistics \cite{tibshirani2001gap} using a uniform reference distribution (denoted as Gap(uni)) and using a principal component reference distribution (denoted as Gap(pc)); the extension of the Gap statistics including Weighted Gap statistics (Wei(uni), Wei(pc)) and DD-Weighted Gap method (DD(uni), DD(pc)) \cite{yan2007ddwei}; the Jump method \cite{sugar2003jump}; spectral clustering Spectrum \cite{john2020spectrum}; and the Gaussian mixture model with criteria BIC and ICL (GMM(BIC), GMM(ICL)) \cite{fraley1998GMM, biernacki2000icl}. 

For the graph-based method, the similarity graph is the 10-MST constructed from Euclidean distance. In simulations that involve resampling, we use 50 bootstraps to create the reference distribution for the Gap statistics and its extensions. The transformation power is set to $5$ in the Jump method. In spectral clustering, the number of clusters is estimated by identifying the biggest gap between eigenvalues obtained through spectral decomposition of the similarity matrix, which is constructed using a self-tuning kernel (see \cite{john2020spectrum} for additional details). 

The data are generated from a $d=400$ dimensional distribution; the specific distributions are provided below in Section \ref{sec:sim_design}. The true number of clusters is $k^\star = 3$ for Scenarios I and III, and is $k^\star = 4$ for Scenarios II, IV, and V. The maximum number of clusters we consider is $k_{max} =10$.

\begin{figure}
\centering
\includegraphics[width=0.8\linewidth]{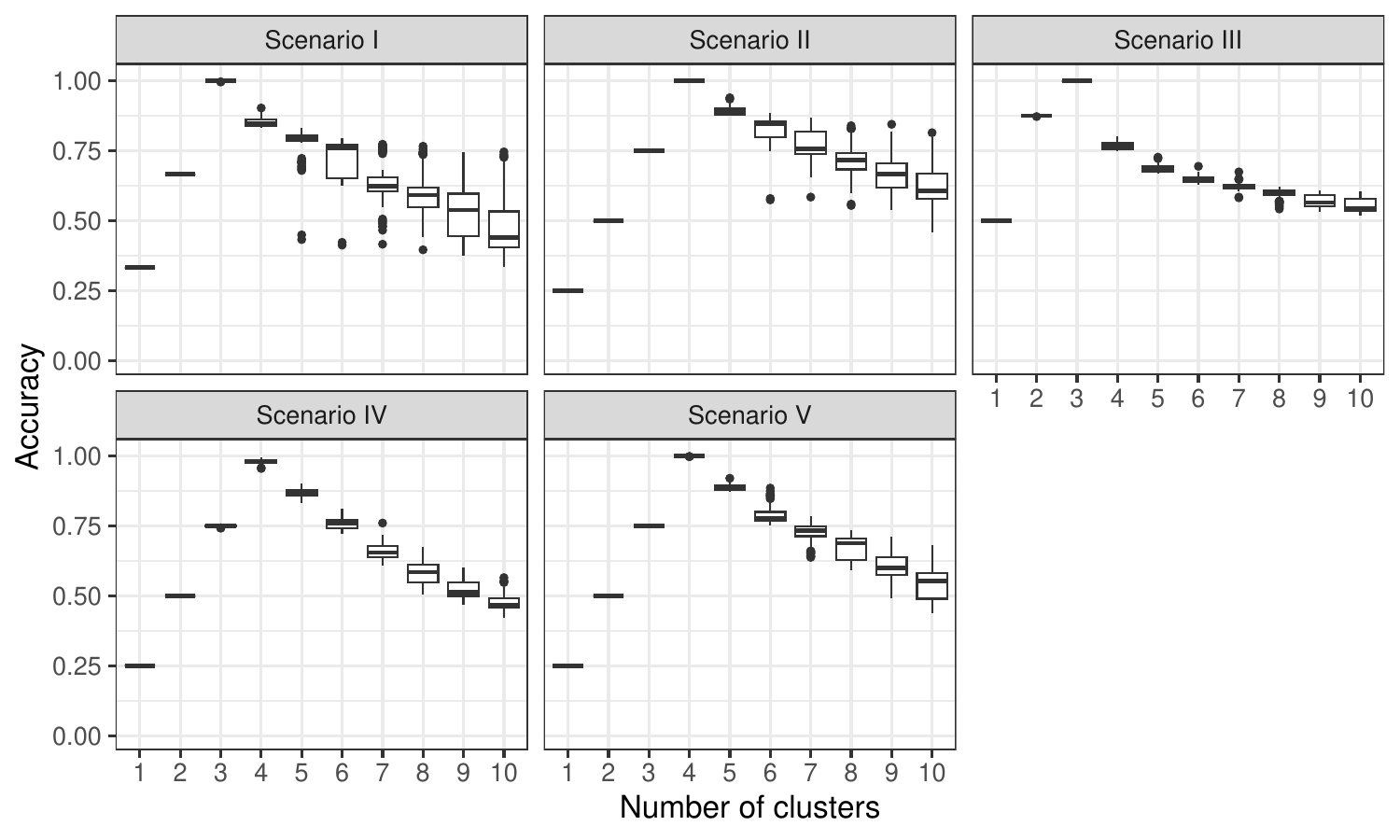}
\caption{Accuracy of the clustering under different numbers of clusters in different scenarios. For Scenarios I and III, $k^\star = 3$. For Scenarios II, IV, and V, $k^\star = 4$.}
\label{fig:acc}
\end{figure}

We utilize $K$-means clustering to perform clustering here. $K$-means clustering can achieve high accuracy under the true number of clusters in our studies. We use the following criteria to assess the clustering accuracy:
$$\text{acc}(c,\hat{c})=\max_{a\in A}\frac{1}{n}\sum_{i=1}^nI(a(\hat{c}_i)=c_i)$$
where $A$ is the set of all permutations for $[1, 2,..., k]$, $a$ is one permutation, $c_i$ is the true label of observation $i$, $\hat{c}_i$ is the estimated label given by $K$-means clustering, and $I$ is the indicator function. In practice, the actual labels are unknown, and the accuracy used here can only be calculated for simulation studies. The clustering accuracies for different numbers of clusters under various scenarios are shown in Figure \ref{fig:acc}; all are  above 95\% under the true number of clusters. 

\subsection{Study Design} \label{sec:sim_design}

We present results for five different simulation scenarios: Scenario I consists of Gaussian data with location and scale differences across clusters; Scenario II consists of observations from Gaussian distributions with one cluster dominating the others; Scenario III has observations with unequal cluster size from Gaussian distributions; Scenario IV involve Gaussian data with correlated features; Scenario V has non-Gaussian data from the Lognormal distribution. The exact simulation settings are given in Supplement Section S1. We perform 100 replicates for each scenario and record the estimated $\hat{k}$ for each method.

\begin{table}[!t]
    \centering  
    \begin{tabular}{@{\extracolsep{\fill}} l| c c c c c c c c c c@{\extracolsep{\fill}}}
        \toprule
        
        $\hat{k}$& 1 & 2 & 3 & 4 & 5 & 6 & 7 & 8 & 9 & 10 \\ 
       \midrule
    Graph-based & - & 0 & \textbf{100} & 0 & 0 & 0 & 0 & 0 & 0 & 0\\  
 TraceW & 0 & 100 & \textbf{0} & 0 & 0 & 0 & 0 & 0 & 0 & 0\\  

 Silhouette & - & 100 & \textbf{0} & 0 & 0 & 0 & 0 & 0 & 0 & 0\\ 
 
 Jump & 0 & 0 & \textbf{0} & 0 & 1 & 8 & 22 & 31 & 38  & 0 \\

 Gap(uni) & 0 & 0 & \textbf{1} & 63 & 32 & 4 & 0 & 0 & 0 & 0\\ 
 Gap(pc)& 0 & 100 & \textbf{0} & 0 & 0 & 0 & 0 & 0 & 0 & 0\\ 

  Wei(uni) & 0 & 0  & \textbf{100} & 0 & 0 & 0 & 0 & 0 & 0 & 0\\ 
  Wei(pc) & 0 & 0 & \textbf{100} & 0 & 0 & 0 & 0 & 0 & 0 & 0\\  

DD(uni) & 0 & 0 & \textbf{78} & 0 & 0 & 1 & 6 & 6 & 6 & 3 \\ 
DD(pc)& 0 & 0 & \textbf{72} & 0 & 0 & 1 & 6 & 12 & 6 & 3 \\  
Spectral& - & 1 & \textbf{99} & 0 & 0 & 0 & 0 & 0 & 0 & 0\\  
  GMM(BIC) & 0 & 100 & \textbf{0} & 0 & 0 & 0 & 0 & 0 & 0 & 0\\ 
  GMM(ICL)& 0 & 100 & \textbf{0} & 0 & 0 & 0 & 0 & 0 & 0 & 0\\  
  \hline
    \end{tabular}
    \caption{The frequency of estimated number of clusters over 100 trials. The results corresponding to the true $k^{\star}$ is bolded for Scenario I.}
    \label{tab:sim1}
\end{table}

\begin{table}[!t]
    \centering  
    \begin{tabular}{@{\extracolsep{\fill}} l| c c c c c c c c c c@{\extracolsep{\fill}}}
        \toprule
        
        $\hat{k}$& 1 & 2 & 3 & 4 & 5 & 6 & 7 & 8 & 9 & 10 \\ 
       \midrule 
       Graph-based & - & 0 & 0 & \textbf{100} & 0 & 0 & 0 & 0 & 0 & 0\\ 
 TraceW & 0 & 100 & 0 & \textbf{0} & 0 & 0 & 0 & 0 & 0 & 0\\  

 Silhouette & - & 100 & 0 & \textbf{0} & 0 & 0 & 0 & 0 & 0 & 0\\ 
 
 Jump & 100 & 0 & 0 & \textbf{0} & 0 & 0 & 0 & 0 & 0 & 0\\

 Gap(uni) & 0 & 0 & 0 & \textbf{5} & 29 & 54 & 10 & 2  & 0 & 0\\ 
 Gap(pc)& 0 & 0 & 0 & \textbf{93} & 6 & 1 & 0 & 0 & 0 & 0\\ 

  Wei(uni) & 0 & 0  & 0 & \textbf{94} & 5 & 0 & 0 & 0 & 0 & 1\\ 
  Wei(pc) & 0 & 0 & 0 & \textbf{94} & 5 & 0 & 0 & 0 & 0 & 1\\  

DD(uni) & 0 & 100 & 0 & \textbf{0} & 0 & 0 & 0 & 0 & 0 & 0 \\ 
DD(pc) & 0 & 100 & 0 & \textbf{0} & 0 & 0 & 0 & 0 & 0 & 0  \\  
Spectral & - & 0 & 0 & \textbf{100} & 0 & 0 & 0 & 0 & 0 & 0\\  
  GMM(BIC) & 0 & 0 & 100 & \textbf{0} & 0 & 0 & 0 & 0 & 0 & 0\\ 
  GMM(ICL)& 0 & 0 & 100 & \textbf{0} & 0 & 0 & 0 & 0 & 0 & 0\\  
  \hline
    \end{tabular}
    \caption{The frequency of estimated number of clusters over 100 trials. The results corresponding to the true $k^{\star}$ is bolded for Scenario II.}
    \label{tab:sim2}
\end{table}

\begin{table}[!t]
    \centering  
    \begin{tabular}{@{\extracolsep{\fill}} l| c c c c c c c c c c@{\extracolsep{\fill}}}
        \toprule
        
        $\hat{k}$& 1 & 2 & 3 & 4 & 5 & 6 & 7 & 8 & 9 & 10 \\ 
       \midrule 
       Graph-based & - & 0 & \textbf{100} & 0 & 0 & 0 & 0 & 0 & 0 & 0\\  

 TraceW & 0 & 100 & \textbf{0} & 0 & 0 & 0 & 0 & 0 & 0 & 0\\  

 Silhouette & - & 0 & \textbf{0} & 0 & 0 & 0 & 2 & 11 & 23 & 64\\ 
 
 Jump & 0 &0 & \textbf{0} & 0 &1 & 2& 19 &33 &45  & 0\\

 Gap(uni) & 0 & 0 & \textbf{0} & 32 & 43 & 18 & 5 & 1 & 0 & 0\\ 
 Gap(pc)& 0 & 0 & \textbf{100} & 0 & 0 & 0 & 0 & 0 & 0 & 0\\ 

  Wei(uni) & 0 & 100  & \textbf{0} & 0 & 0 & 0 & 0 & 0 & 0 & 0\\ 
  Wei(pc) & 0 & 100 & \textbf{0} & 0 & 0 & 0 & 0 & 0 & 0 & 0\\  

DD(uni) & 0 & 100  & \textbf{0} & 0 & 0 & 0 & 0 & 0 & 0 & 0\\ 
DD(pc)& 0 & 100  & \textbf{0} & 0 & 0 & 0 & 0 & 0 & 0 & 0\\  
Spectral & - & 0 & \textbf{100} & 0 & 0 & 0 & 0 & 0 & 0 & 0\\  
  GMM(BIC) & 0 & 0 & \textbf{100} & 0 & 0 & 0 & 0 & 0 & 0 & 0\\ 
  GMM(ICL)& 0 & 0 & \textbf{100} & 0 & 0 & 0 & 0 & 0 & 0 & 0\\  
  \hline
    \end{tabular}
    \caption{The frequency of estimated number of clusters over 100 trials. The results corresponding to the true $k^{\star}$ is bolded for Scenario III.}
    \label{tab:sim3}
\end{table}

\begin{table}[!t]
    \centering  
    \begin{tabular}{@{\extracolsep{\fill}} l| c c c c c c c c c c@{\extracolsep{\fill}}}
        \toprule
        
        $\hat{k}$& 1 & 2 & 3 & 4 & 5 & 6 & 7 & 8 & 9 & 10 \\ 
       \midrule 
       Graph-based & - & 0 & 0 & \textbf{100} & 0 & 0 & 0 & 0 & 0 & 0\\  

TraceW & 0 & 100 & 0 & \textbf{0} & 0 & 0 & 0 & 0 & 0 & 0\\  

 Silhouette & - & 100 & 0 & \textbf{0} & 0 & 0 & 0 & 0 & 0 & 0\\ 
 
  Jump & 100 &0 & 0 & \textbf{0} & 0 & 0 & 0 & 0 & 0 & 0\\
     Gap(uni) & 0 & 0 & 0 & \textbf{100} & 0 & 0 & 0 & 0 & 0 & 0\\ 
 Gap(pc)& 0 &100 & 0 & \textbf{0} & 0 & 0 & 0 & 0 & 0 & 0\\ 

  Wei(uni) & 0 & 0  & 0 & \textbf{69} & 28 & 2 & 1 & 0 & 0 & 0\\ 
  Wei(pc) & 0 & 100 & 0 & \textbf{0} & 0 & 0 & 0 & 0 & 0 & 0\\  

DD(uni) & 0 & 100  & 0 & \textbf{0} & 0 & 0 & 0 & 0 & 0 & 0\\ 
DD(pc)& 0 & 100  & 0 &\textbf{0} & 0 & 0 & 0 & 0 & 0 & 0\\  
Spectral & - & 97 & 3 & \textbf{0} & 0 & 0 & 0 & 0 & 0 & 0\\  
  GMM(BIC) & 0 & 0 & 100 & \textbf{0} & 0 & 0 & 0 & 0 & 0 & 0\\ 
  GMM(ICL)& 0 & 0 & 100 & \textbf{0} & 0 & 0 & 0 & 0 & 0 & 0\\  
  \hline
    \end{tabular}
    \caption{The frequency of estimated number of clusters over 100 trials. The results corresponding to the true $k^{\star}$ is bolded for Scenario IV.}
    \label{tab:sim4}
\end{table}

\begin{table}[!t]
    \centering  
    \begin{tabular}{@{\extracolsep{\fill}} l| c c c c c c c c c c@{\extracolsep{\fill}}}
        \toprule
        
        $\hat{k}$& 1 & 2 & 3 & 4 & 5 & 6 & 7 & 8 & 9 & 10 \\ 
       \midrule 
       Graph-based & - & 0 & 0 & \textbf{100} & 0 & 0 & 0 & 0 & 0 & 0\\  

 TraceW & 0 & 100 & 0 & \textbf{0} & 0 & 0 & 0 & 0 & 0 & 0\\  

 Silhouette & - & 100 & 0 & \textbf{0} & 0 & 0 & 0 & 0 & 0 & 0\\ 
 
 Jump & 100 &0 & 0 & \textbf{0} & 0 & 0 & 0 & 0 & 0 & 0\\

 Gap(uni) & 0 & 0 & 0 & \textbf{3} & 7 &22 & 31 & 19 & 13 & 0\\ 
 Gap(pc)& 0 & 0 & 0 & \textbf{100} & 0 & 0 & 0 & 0 & 0 & 0\\ 

  Wei(uni) & 0 & 0 & 0 & \textbf{99} & 1 & 0 & 0 & 0 & 0 & 0\\ 
  Wei(pc) & 0 & 0 & 0 & \textbf{99} & 1 & 0 & 0 & 0 & 0 & 0\\  

DD(uni) & 0 & 100 & 0 &\textbf{0} & 0 & 0 & 0 & 0 & 0 & 0\\ 
DD(pc)& 0 & 100 & 0 & \textbf{0} & 0 & 0 & 0 & 0 & 0 & 0\\  
Spectral & - & 0 & 14 & \textbf{86} & 0 & 0 & 0 & 0 & 0 & 0\\  
  GMM(BIC) & 0 & 0 & 66 & \textbf{11} & 6 & 3 & 3 & 6 & 3 & 2\\ 
  GMM(ICL)& 0 & 0 & 66 & \textbf{11} & 6 & 3 & 3 & 6 & 3 & 2\\
  \hline
    \end{tabular}
    \caption{The frequency of estimated number of clusters over 100 trials. The results corresponding to the true $k^{\star}$ is bolded for Scenario V.}
    \label{tab:sim5}
\end{table}

Tables \ref{tab:sim1} -  \ref{tab:sim5} present the simulation results and report the frequency with which a specific value of $\hat{k}$ was estimated across 100 experiments in each of the five scenarios. The results under the true $k^\star$ are bolded. We can see that using $W_k$ or $B_k$ as in `TraceW', `Silhouette', and `Jump' does not work well in high-dimensional settings. 

For the remaining methods, each have specific scenarios where they perform well, but none can consistently outcompete the graph-based model across all settings. For example, the Gap statistic utilizing the uniform reference distribution exhibits limited effectiveness in Scenarios I - III and V when the difference between clusters goes beyond mean change. It may fail to provide an estimated value when the data is non-Gaussian (note that the total number of trials for Gap(uni) in Scenario V does not sum to 100). Gap statistics using a principal component reference perform poorly when the dimensions are correlated (Scenario IV); the method tends to underestimate the number of clusters. The weighted gap and the DD-weighted gap show slight performance gains compared to the Gap statistics under most of the settings. On the other hand, spectral clustering performs well in Scenarios I-III, but confronts difficulties when the data have correlated features (Scenario IV) or follow a non-Gaussian distribution with a long tail (Scenario V). With the exception of Scenario III, both GMM methods tend to pick simpler models with $\hat{k} < k^{\star}$. 

The discerning reader may note that $K$-means is also a distance-based procedure and appears to perform quite well even in this high-dimensional setting. In our simulation studies, the clusters are well-separated. Since $K$-means focuses on an optimal local structure, such that observations with smaller distances are grouped together for a fixed value of $k$, the \textit{local structure based on the ordering of distances} can still effectively partition the observations. However, we can see that existing distance-based methods, such as TraceW, are based on comparing global distance information (such as the total within-cluster dispersion) across different values of $k$. These global metrics are not as informative in high-dimensional space, which limits their ability to effectively estimate $k$ consistently across different simulation scenarios.  

\section{Real Data Application}\label{sec:app}
\subsection{Fashion-MNIST image data}

\begin{figure}[!ht]
\centering
\includegraphics[width=0.8\linewidth]{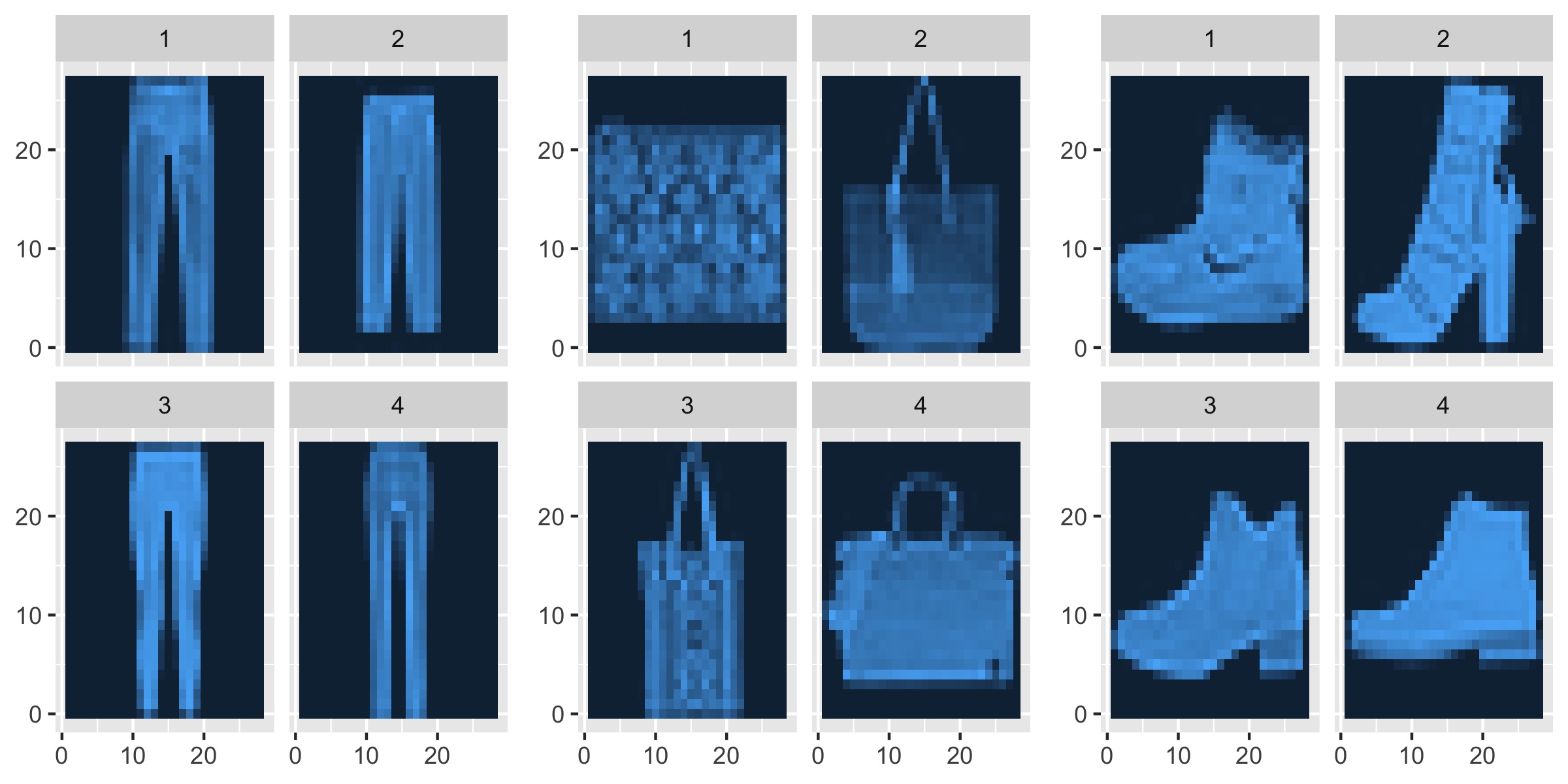}
\caption{Sample fashion images of trousers, bags, and ankle boots.}
\label{fig:fashion}
\end{figure}

We evaluate our method on the Fashion-MNIST image dataset \cite{xiao2017fashion}. This is a dataset of Zalando's article images; it consists of a training set of 60,000 observations and a test set of 10,000 observations. Each observation is a grayscale image centered in a $28 \times 28$ pixel box associated with a label from one of 10 classes. To illustrate our approach, we select the categories trouser, bag, and ankle boot from the original dataset so that $k^\star = 3$. Here, we know the true labels of the images so we can directly assess the methods' performance. Examples of the images are shown in Figure \ref{fig:fashion}. The high-dimensional nature of the pixel data makes estimating the number of clusters inherently difficult. We run 100 experiments, and for each experiment, we randomly select 500 images from the training set to perform clustering. 

We compare our approach to a spectral clustering approach proposed in \cite{john2020spectrum}. For the graph-based approach, the similarity graph is the 80-MST constructed from the same similarity matrix as the one in spectral clustering. The estimation results are shown in Table \ref{tab:app1}. Compared to the spectral clustering approach, we can see that the graph-based method demonstrates better performance in this application for estimating $k$.
        
\begin{table}[!ht]
    \centering
    
    \begin{tabular}{@{\extracolsep{\fill}}l cc@{\extracolsep{\fill}}}
        \toprule
        
       Method& Graph-based & Spectral\\
       \midrule
      Number of trials with $\hat{k}=3$ &93&64  \\
        \bottomrule
    \end{tabular}
    \caption{Fashion-MNIST estimation results from 100 trials comparing the graph-based method and the spectral clustering approach. The true number of clusters is 3.}
    \label{tab:app1}
\end{table}

\subsection{Performance on RNA-seq data}
\begin{figure}[!ht]
\centering
\includegraphics[width=\linewidth]{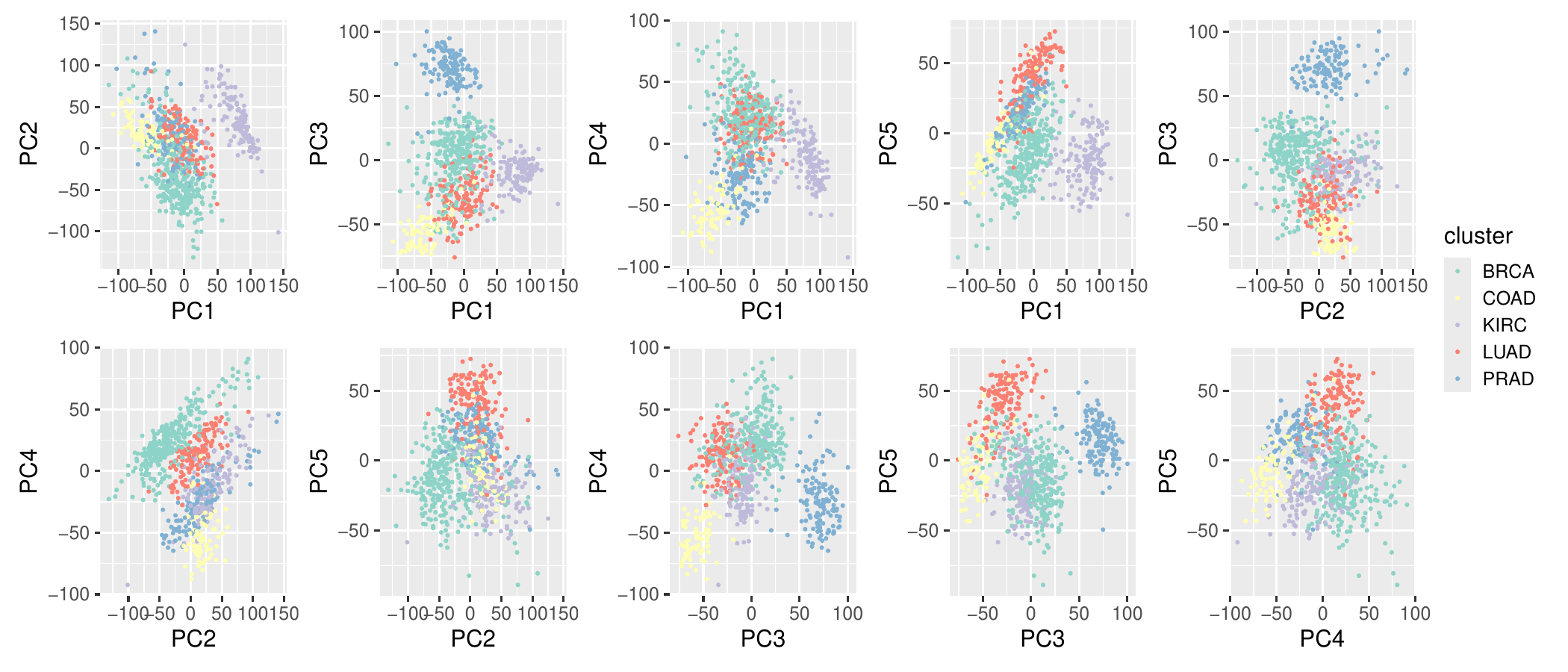}
\caption{Pairwise scatter plots of the first five principal components for the cancer RNA-Seq dataset.}
\label{fig:rna}
\end{figure}

We also apply our method to the UCI gene expression cancer RNA-Seq dataset from the UCI repository \cite{gene}, which consists of 801 patients with 20,531 genes. The data originated from the Atlas Pan-Cancer project with RNA information for five different types of tumors: lung adenocarcinoma (LUAD), breast carcinoma (BRCA), kidney renal clear-cell carcinoma (KIRC), colon adenocarcinoma (COAD), and Prostate adenocarcinoma (PRAD) \cite{weinstein2013cancer}. Pairwise scatter plots of the first five principal components are shown in Figure \ref{fig:rna}. The tumor clusters are not apparent by just observing the plots in low dimensions. We do not use the tumor label information and evaluate the performance of different methods to estimate the number of clusters for this RNA dataset.

We compare the graph-based method with TraceW, Silhouette, Gap statistics, Weighted Gap statistics, DD-Weighted Gap statistics, and the Jump method. We conduct principal component analysis and use $K$-means clustering on the first 200 principal components; 81\% of the variance is explained by the first 200 PC scores. 
The competing methods are applied to the 200 PC scores to estimate $k$, and, when possible, to the full dataset of 20,531 genes. In the latter setting, the results of Gap statistics, weighted Gap, and DD-weighted Gap statistics are not reported since these methods were unable to estimate the number of clusters. For the graph-based method, we construct a 50-MST in both settings.  The results are shown in Table \ref{tab:app2}. Using PC scores, the graph-based method and both weighted Gap statistics can correctly recognize the true number of clusters, while the other methods encounter obstacles in identifying the true $k^\star$. In the full data setting, only the graph-based approach successfully estimates $k^\star$. 

\begin{table}[!ht]
    \centering
    \begin{tabular}{@{\extracolsep{\fill}}l cc@{\extracolsep{\fill}}}
        \toprule
        & \multicolumn{2}{c}{$\hat{k}$}\\
        \cline{2-3}
        Method & principal components & full data\\
        \hline
        Graph-based &\textbf{5}& \textbf{5}\\
        TraceW &2 & 2\\
        Silhouette &7 & 6\\
        Gap(uni) &10 & --\\
        Gap(pc)&10& --\\
        WeiGap(uni) &\textbf{5}& --\\
        WeiGap(pc) &\textbf{5}& --\\
        DD-WeiGap(uni) &2& --\\
        DD-WeiGap(pc)&2& --\\
        Jump&4 & 4\\
        \bottomrule
    \end{tabular}
    \caption{Estimated number of clusters for different methods using 200 PC scores (principal components) and the full dataset.}
    \label{tab:app2}
\end{table}

\section{Discussion}\label{sec:dis}

\subsection{Checking the assumption for asymptotic consistency}\label{dis:ass}
In practice, verifying Assumption \ref{as:1} is difficult due to incomplete knowledge regarding the true density functions. Moreover, the integration over densities is challenging to obtain when the dimension of the observations is moderately large. The assumption can be checked under certain conditions. For example, when the observations in each cluster come from densities with disjoint supports, Assumption \ref{as:1} always holds. The proof is given in Supplement Section S9
. In this section, we explore some simple settings to check the assumption. 

\subsubsection{Gaussian examples ($d=2$)}
We present some examples to illustrate validating the assumption when the true densities of the clusters are Gaussian. 
In order for the integral in Assumption \ref{as:1} to be tractable, we consider the low-dimensional case when the dimension of each observation is two. 

\noindent \textbf{Example 1:}
We simulate four clusters that differ in means: $f_1 \sim \mathcal{N}((0,$ $0)^T, I_2)$, $f_2 \sim \mathcal{N}((4, 0)^T, I_2)$, $f_3 \sim \mathcal{N}((4, 4)^T, I_2)$ and $f_4 \sim \mathcal{N}((0, 4)^T, I_2)$.

Let $\widetilde{\mathcal{F}} = \{f_1, f_2, f_3, f_4\}$ be the true set of all densities, and $p_j = 0.25, \forall j \in \{1, 2, 3, 4\}$. 
Then we can calculate the pairwise divergences $\mathcal{D}$ between densities as follows:
\begin{align*}
&\mathcal{D}(f_1, f_2|\widetilde{\mathcal{F}}) = \mathcal{D}(f_1, f_4|\widetilde{\mathcal{F}}) =   \mathcal{D}(f_2, f_3|\widetilde{\mathcal{F}}) =\mathcal{D}(f_3, f_4|\widetilde{\mathcal{F}}) = 0.4497283, \\
&\mathcal{D}(f_1, f_3|\widetilde{\mathcal{F}}) = \mathcal{D}(f_2, f_4|\widetilde{\mathcal{F}}) = 0.4657013.
\end{align*}
The true number of clustering is $k^\star = 4$. Since the divergence between $f_1$ and $f_2$ is among the smallest, we examine a cluster assignment that contains a mixture cluster of $f_1$ and $f_2$. Specifically, consider a cluster assignment, $\psi_3$, that assigns the observations to three clusters, such that $C_1$ consists of observations from $f_1$ and $f_2$, $C_2$ consists of observations from $f_3$, and $C_3$ consists of observations from $f_4$. Under this assignment $\psi_3$, Assumption \ref{as:1} boils down to checking the following inequality:
$$\frac{1}{0.5}\mathcal{D}(f_{1}, f_{2}|\widetilde{\mathcal{F}})
        >\left(1-\frac{\sqrt{3-1}}{\sqrt{3+1-1}}\right)\sum_{i = 1}^4\sum_{j=1, j>i}^4\mathcal{D}(f_{i}, f_{j}|\widetilde{\mathcal{F}}),$$
since $H' = \{1\}$, $|H'| =1$, $m_1 = 2$, and $k^+=1$. 
Since the left-hand side is $0.8994567$ which is larger than the right-hand side $0.5010223$, Assumption \ref{as:1} holds.

Alternatively, consider a clustering assignment $\psi_2$, such that observations from $f_1$ and $f_2$ are assigned to $C_1$ and observations from $f_3$ and $f_4$ are assigned to $C_2$.  Then Assumption \ref{as:1} involves checking the following inequality: \begin{align*}
    &\frac{1}{0.5}\mathcal{D}(f_{1}, f_{2}|\widetilde{\mathcal{F}}) + \frac{1}{0.5}\mathcal{D}(f_{3}, f_{4}|\widetilde{\mathcal{F}}) >\left(1-\frac{\sqrt{2-1}}{\sqrt{2+2-1}}\right)\sum_{i = 1}^4\sum_{j=1, j>i}^4\mathcal{D}(f_{i}, f_{j}|\widetilde{\mathcal{F}}),
\end{align*} 
        since $H' = \{1, 2\}$, $|H'| =2$, $m_1 = 2, m_2 = 2$, and $k^+=2$. 
The left-hand side of the inequality is $1.798913$ and the right-hand side is $1.153967$. Again, the assumption holds.  

\textbf{Example 2:}
Now we consider scenarios where assumption is violated. This occurs if the mixture clusters contain densities that are similar in comparison to the total divergence. Consider the four following clusters with only the mean differences: $f_1 \sim N((0, 0), I_2)$, $f_2 \sim N((1, 0), I_2)$, $f_3 \sim N((5, 5), I_2)$ and $f_4 \sim N((0, 5), I_2)$.
  
Given the set of true densities $\widetilde{\mathcal{F}} = \{f_1, f_2, f_3, f_4\}$, and $p_j = 0.25, \forall j \in \{1, 2, 3, 4\}$, the pairwise divergences $\mathcal{D}$ are as follows:
\begin{align*}
    &\mathcal{D}(f_1, f_2|\widetilde{\mathcal{F}}) = 0.1012585,
    \mathcal{D}(f_1, f_3|\widetilde{\mathcal{F}}) = 0.3967714,
    \mathcal{D}(f_1, f_4|\widetilde{\mathcal{F}}) = 0.3902628,\\
    &\mathcal{D}(f_2, f_3|\widetilde{\mathcal{F}}) = 0.3966169,
    \mathcal{D}(f_2, f_4|\widetilde{\mathcal{F}}) = 0.3913029,
    \mathcal{D}(f_3, f_4|\widetilde{\mathcal{F}}) = 0.4872810.
\end{align*}
Consider a cluster assignment where $f_1$ and $f_2$ are assigned to the same cluster $C_1$, observations from $f_3$ are assigned to $C_2$, and observations from $f_4$ are assigned to $C_3$. Since $H' = \{1\}$, $|H'| =1$, $m_1 = 2$, and $k^+=1$, we need to check the following inequality:
 $$\frac{1}{0.5}\mathcal{D}(f_{1}, f_{2}|\widetilde{\mathcal{F}})
        >\left(1-\frac{\sqrt{3-1}}{\sqrt{3+1-1}}\right)\sum_{i = 1}^4\sum_{j=1, j>i}^4\mathcal{D}(f_{i}, f_{j}|\widetilde{\mathcal{F}}).$$ The assumption does not hold since the left-hand side of the inequality (0.202517) is smaller than the right-hand side (0.3970085). This happens because the mixture cluster, which contains observations from $f_1$ and $f_2$, is not well separated enough relative to the other pairwise distances. In fact, we can see that they are the most similar in terms of divergence in the set $\widetilde{\mathcal{F}}$.

Additional examples can be found in Supplement Section S10 
and include settings where the clusters are unbalanced or the densities differ in both mean and variance.

\subsection{Choice of $\mathcal{K}$}\label{choi:k}

\begin{figure}[!t]
\centering
\includegraphics[width=1\linewidth]{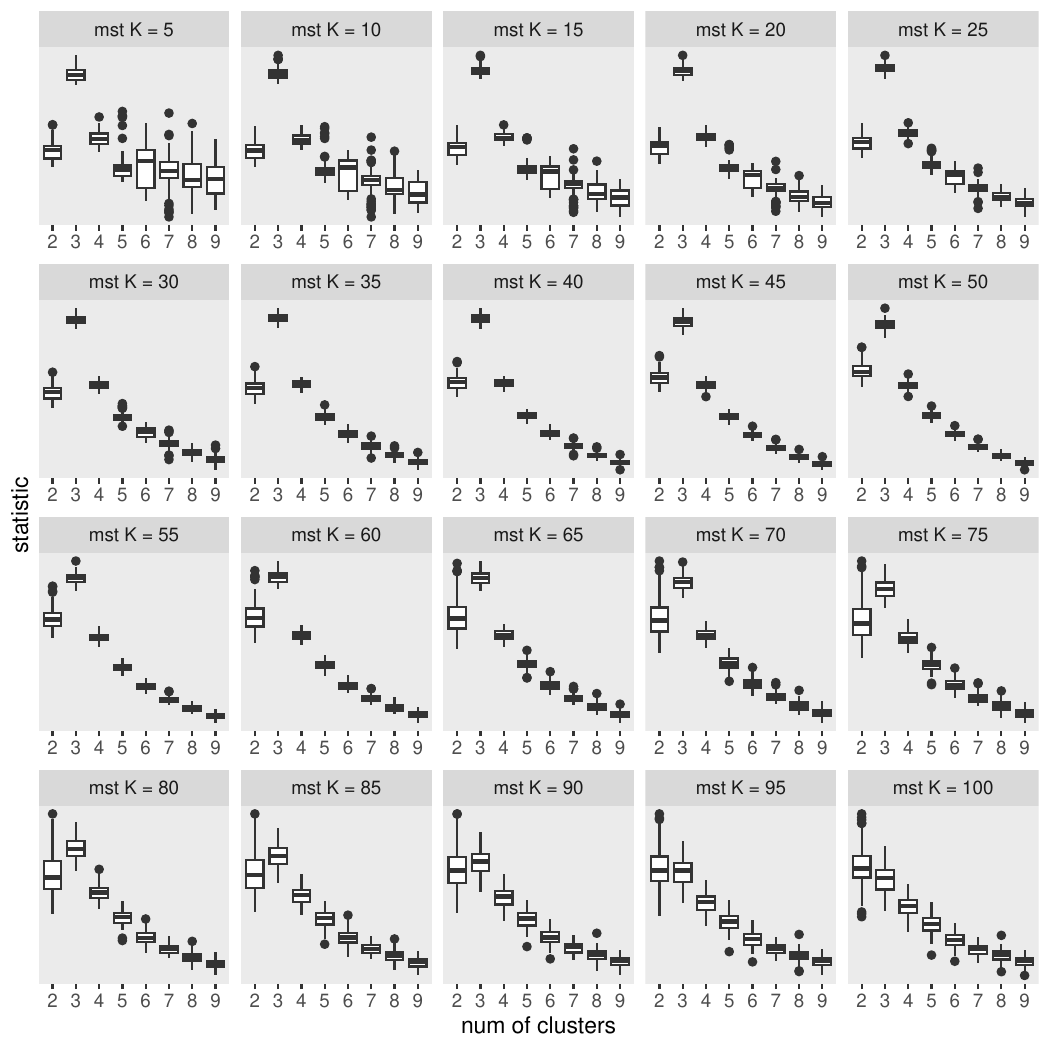}
\caption{Boxplots of the graph-based statistics over 100 simulations for different values of $\mathcal{K}$ in $\mathcal{K}$-MST.}
\label{fig:mst_k_inc}
\end{figure}

The density of the similarity graph is controlled by the choice of $\mathcal{K}$ in a $\mathcal{K}$-MST, and similarly for a $\mathcal{K}$-NN. If $\mathcal{K}$ is too small, it is possible insufficient similarity information will be captured by the graph which could result in a less informative statistic. However, a denser similarity graph will not always yield superior results since, as $\mathcal{K}$ increases, less relevant similarity information may be incorporated into the statistic. To demonstrate the effect of $\mathcal{K}$ on the performance of our method to estimate $k$, we simulate data from the following densities:

\begin{itemize}
    \item  $C_1 \sim \mathcal{N} (0\cdot\mathbf{1}_{400}, \mathbf{I}_{400}), n_1  =100$,
    \item  $C_2 \sim \mathcal{N} ([\begin{smallmatrix}1.5\cdot\mathbf{1}_{200}, 0\cdot\mathbf{1}_{200}\end{smallmatrix}], [\begin{smallmatrix} 1.3\cdot\mathbf{I}_{200} & \\ & \mathbf{I}_{200}\end{smallmatrix}]), n_2  =100$,
    \item  $C_3 \sim \mathcal{N} ([\begin{smallmatrix}0.8\cdot\mathbf{1}_{200}, 0\cdot\mathbf{1}_{200}\end{smallmatrix}], [\begin{smallmatrix} 1.5\cdot\mathbf{I}_{200} & \\ & \mathbf{I}_{200}\end{smallmatrix}]), n_3  =100$.
\end{itemize}
Figure \ref{fig:mst_k_inc} presents boxplots of the graph-based statistics for various values of $k$ across 100 experiments as $\mathcal{K}$ increases.

For values of $\mathcal{K}$ from 5 to 55, we see that the graph-based method can correctly estimate $k^\star=3$, with the variability improving as $k$ increases. However as $\mathcal{K}$ goes beyond 55, potentially redundant or less relevant similarity information is incorporated into the statistic. We observe that the method tends to select a smaller number of clusters and this becomes more extreme as $\mathcal{K}$ increases to 100, to the point where the graph-based statistic estimates $\hat{k}=2$. Thus, in practice, $\mathcal{K}$ must be chosen carefully and a range of values can be initially considered. Based on our simulations, we recommend $\mathcal{K}$ = 30 as a starting point.

\section{Conclusion}\label{sec:conclusion}

We propose a graph-based approach for estimating the number of clusters. Our approach is computationally efficient, supported by asymptotic theory, performs well for observations in arbitrary dimensions, and can be paired alongside any clustering algorithm. The statistic is constructed from the (standardized) within-cluster edge count of a graph, which is an informative metric in distinguishing similarity within and between clusters. The proposed estimator $\hat{k}$ is the value of $k$ that maximizes the graph-based statistic. We establish that the statistic is maximized at the true $k^\star$, corresponding to $\widetilde{\psi}_{k^\star}$, and prove the selection consistency of the estimator. This ensures that the number of clusters selected will converge in probability to the true number of clusters $k^\star$, assuming that the densities within the mixture clusters are separated to some extent. The consistency is established by showing that for any clustering other than the true clustering, we can always find another partition of the observations that generates a larger value of the statistic as $n\rightarrow \infty$. We evaluate the performance of the graph-based approach via simulation studies and real data applications by comparing it to other commonly used methods to estimate $k$. We see that the graph-based approach can consistently perform well in a range of scenarios when the dimension of the observations is moderate-to-high. Many avenues for future research can be explored. For example, the condition in Theorem \ref{th:main} could potentially be relaxed. Extensions to directed or weighted graphs could yield a more informative graph-based statistic. Furthermore, this framework could also be adapted to other settings, such as estimating the number of communities in a network.  

\bibliographystyle{plain}
\bibliography{reference}

\end{document}